\documentclass[preprint,showpacs,preprintnumbers,amsmath,amssymb,nofootinbib]{revtex4}

\usepackage{etex}
\usepackage{graphicx}
\usepackage{dcolumn}
\usepackage{bm}
\usepackage{amssymb,amsthm,amscd,amsbsy,array}\usepackage{amsmath,dsfont}

\usepackage{graphics,xcolor}

\usepackage{color}

\newcommand{\SU}{\mathrm{SU}}

\newcommand{\half}{{\scriptstyle{\frac{1}{2}}}}
\def\2{{\half}}
\newcommand{\const}{\mathop{\rm const}\nolimits}

\def\bA{{\bm{A}}}

\def\su{{\mathrm{su}}}
\def\p{{\partial}}

\def\bomega{{\bm{\omega}}}

\def\bOmega{\mbox{\boldmath$\omega$}}

\def\br{{\bm{r}}}

\def\hx{{\hat{\br}}}

\def\beq{\begin{equation}}
\def\eeq{\end{equation}}
\def\beqa{\begin{eqnarray}}
\def\eeqa{\end{eqnarray}}

\def\barray{\left(\begin{array}}
\def\earray{\end{array}\right)}
\def\barraynb{\begin{array}}
\def\earraynb{\end{array}}

\def\IS{{\mathds{S}}}

\def\smallover#1/#2{\hbox{$\textstyle\frac{#1}{#2}$}} %
\def\vI{{\vec{Q}}}
\def\vW{{\vec{\Psi}}}
\def\vN{{\vec{N}}}
\def\vK{{\vec{K}}}
\def\vD{{\vec{D}}}
\def\vJ{{\vec{J}}}
\def\vpi{{\vec{\pi}}}
\def\vx{{\br}}
\def\vn{{\vec{n}}}
\def\vp{{\vec{p}}}

\def\vA{{\vec{A}}}

\def\vC{{\vec{C}}}

\def\tomega{{\tilde{\omega}}}
\def\tOmega{{\tilde{\Omega}}}
\newcommand{\kernel}{\mathrm{Ker}\,}
\newcommand{\tr}{\mathrm{tr}}

\newcommand{\fG}{\mathfrak{G}}
\newcommand{\fS}{\mathfrak{S}}
\newcommand{\IP}{\mathfrak{P}} 

\newcommand{\cV}{\mathcal{V}}
\newcommand{\cU}{\mathcal{U}}

\newcommand{\cO}{{\mathcal{O}}}

\def\smallcirc{{\raise 0.5pt \hbox{$\scriptstyle\circ$}}}
\def\aand{{\quad\text{\small and}\quad}}
\def\where{{\quad\text{\small where}\quad}}
\def\with{{\quad\text{\small with}\quad}}
\def\ie{{\;\text{\small i.e.}\;}}
\def\ie,{{\;\text{\small i.e.,}\;}}

\def\benu{\begin{enumerate}}
\def\eenu{\end{enumerate}}
\def\bitem{\begin{itemize}}
\def\eitem{\end{itemize}}

\def\besub{\begin{subequations}}
\def\esub{\end{subequations}}

\def\?{{\,\gb{\fbox{\texttt{??}}\;}}\,}

\def\Rarrow{{\quad\Rightarrow\quad}}

\usepackage[colorlinks=true, pdfstartview=FitV, linkcolor=blue, citecolor=blue, urlcolor=blue]{hyperref} 
\newcommand{\cyan}{\textcolor{cyan}}
\newcommand{\magenta}{\textcolor{magenta}}

\newcommand{\orange}{\textcolor{orange}}

\newcommand{\gb}{\quad\colorbox{green}}

\newcommand{\dgreen}{\textcolor[rgb]{0,0.5,0}}

\newenvironment{redtext}{\color{red}}
{\ignorespacesafterend}
\newenvironment{bluetext}{\color{blue}}{\ignorespacesafterend}
\newenvironment{greentext}{\color{green}}{\ignorespacesafterend}
\newenvironment{magentatext}{\color{magenta}}{\ignorespacesafterend}
\newenvironment{cyantext}{\color{cyan}}{\ignorespacesafterend}
\newenvironment{orangetext}{\color{orange}}
{\ignorespacesafterend}

\newcommand{\bmagenta}{\begin{magentatext}}
\newcommand{\emagenta}{\end{magentatext}}
\newcommand{\bcyan}{\begin{cyantext}}
\newcommand{\ecyan}{\end{cyantext}}
\newcommand{\bblue}{\begin{bluetext}}
\newcommand{\eblue}{\end{bluetext}}
\newcommand{\bred}{\begin{redtext}}
\newcommand{\ered}{\end{redtext}}

\newcommand{\bgreen}{\begin{greentext}}
\newcommand{\egreen}{\end{greentext}}
\newcommand{\borange}{\begin{orangetext}}
\newcommand{\eorange}{\end{orangetext}}

\numberwithin{equation}{section}

\let\ssection=\section
\renewcommand{\section}{\setcounter{equation}{0}\ssection}

\newcommand{\bigbox}[1]{\fbox{%
\rule[-20pt]{0pt}{45pt}$\;\;\displaystyle{#1}\;\;$}
}
\newcommand{\medbox}[1]{\fbox{%
\rule[-10pt]{0pt}{25pt}$\;\;\displaystyle{#1}\;\;$}%
}

\newcommand{\NA}{non-Abelian \:}

\begin{document}

\preprint{arXiv:2310.19715v4 [math-ph]}

\title{Kerner equation for motion in a non-Abelian gauge field\footnote{To be published in : {\sl The Languages of Physics - A Themed Issue in Honor of Professor Richard 
Kerner on the Occasion of His 80th Birthday}. Special volume of ``Universe'' (2023). }}

\author{
P. A. Horv\'athy$^{1}$\footnote{mailto:horvathy@univ-tours.fr}
and 
P.-M. Zhang$^{2}$\footnote{ mailto:zhangpm5@mail.sysu.edu.cn},
}

\affiliation{
$^1$ Institut Denis Poisson CNRS/UMR 7013 - Universit\'e de Tours - Universit\'e d'Orl\'eans Parc de Grandmont, 37200, Tours, (France)
\\
$^2$ School of Physics and Astronomy, Sun Yat-sen University, Zhuhai 519082, (China)
\\
}
\date{\today}

\begin{abstract}
The equations of motion of an isospin-carrying particle in a Yang-Mills and gravitational field were first proposed in 1968 by Kerner, who considered geodesics in a Kaluza-Klein-type framework.  Two years later the flat space Kerner equations were completed by considering also the motion of the isospin by Wong, who used a  field-theoretical approach.
Their groundbreaking work was then  followed by a long series of rediscoveries whose history is reviewed. The concept of isospin charge and the physical meaning of its motion are discussed. Conserved  quantities are studied for Wu-Yang monopoles and for diatomic molecules by using van Holten's algorithm.
\\

Universe \textbf{9} (2023) no.12, 519
doi:10.3390/universe9120519
[arXiv:2310.19715 [math-ph]].
\end{abstract}


\maketitle

\tableofcontents

\section{Introduction:  a short history of the isospin 
}\label{Intro}

Certain ideas are put forward, then forgotten, and then reproposed again by various authors who ignore previous, and indeed each other's work.  
A typical example is that of an \emph{isospin-carrying particle moving in a Yang-Mills field}, first studied by Kerner \cite{Kerner68}~:
\begin{figure}[h]\vskip-5mm
\includegraphics[scale=.7]{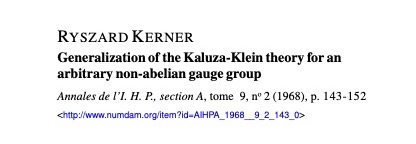}\\{}\vskip-8mm
\caption: {\textit{\small Kerner's paper in which the equation for a particle in a Non-Abelian gauge field was proposed}. 
\label{KernerpKK}
}
\vskip-3mm
\end{figure}

Two years after Kerner's pioneering paper  Wong \cite{Wong70}, who ignored  all about Kerner' work and used a different, field-theoretical  framework, completed the Kerner equations \eqref{Kernereq} below   
which describe motion in ordinary space-time,  with one for the dynamics of the isospin, eqn. \eqref{isospineq}. 

 Their  work was subsequently continued by many other researchers \cite{Trautman70,Cho75,Balachandran76,Balachandran77,Sternberg77,Sternberg78,Weinstein78,Sternberg80,DuvalCRAS,DuvalAix79,DH82, Montgomery,FeherAPH}.
  Jackiw and Manton \cite{JackiwManton}, searching for a physical interpretation of some  quantities found in the study of the symmetries of gauge fields
(re)covered the Kerner equation \eqref{Kernereq}  from a partial variational principle, while {assuming} the isospin-equation \eqref{isospineq}.
 
These studies were parallelled by physical applications 
which include motion in the field of a non-Abelian monopole \cite{GoddardOlive}, which requires extension to Yang-Mills-Higgs systems \cite{Feher:1984ik,Feher:1984xc,Feher86,Feher:1988th}.

Yet another application is to the \emph{non-Abelian Aharonov experiment} proposed by Wu and Yang \cite{WuYang75,WY76}, elaborated in \cite{HPNABA} 
\footnote{To study the Non-Abelian Aharonov-Bohm effect was suggested to one of us (PAH)  in the early eighties by Tai Tsun Wu, who also insisted that we should study the original paper of Yang and Mills \cite{YangMills}. We are grateful for his advices and would like to congratulate also him on his 90th birthday.
} which will be further studied in  \cite{EZH-NABA}. 
The effect is related to topological defects \cite{Alford90,Preskill90,Brandenberger93} and more recently, it  to artificial gauge fields which can be produced in laboratory   
\cite{Osterloh,Dalibard,Goldman,Jacob,YChen,YYang,YBiao,Cserti}. 

As physical illustration, we derive conserved  quantities for Wu-Yang monopoles \cite{WuYang69} and for diatomic molecules \cite{MSW,Jackiw86}. 

\smallskip   
This review  celebrates the 80th birthday of Richard Kerner by recounting the fascinating story of isospin-carrying particles initiated by him when his given name was still ``Ryszard''. 
\goodbreak

\section{Gauge theory and the Kaluza-Klein framework}
\label{Kernerfw}

\subsection{Yang-Mills theory}\label{YMSec} 

The concept of isotopic spin (in short: isospin) was introduced by Heisenberg in 1932 \cite{Heisenberg32}, who argued that a proton and a neutron should be viewed as two different states of the same particle, related by an ``internal'' $\SU(2)$ rotation
\footnote{The fascinating story of gauge theory is recounted by O'Raifeartaigh  \cite{Lochlainn}.}. 

Let us recall that electrodynamics is an Abelian gauge theory: it is described by a real 1-form  $A =A_{\mu} dx^\mu$ called the vector potential which is however determined only up to a gauge transformation,
\beq
A_{\mu} \to A_{\mu} - i{}g^{-1} \p_{\mu}g\,,
\label{U1gauge}
\eeq 
where $g(x^{\mu})$ is an ${\rm U}(1)$-valued function on space-time. 

Twenty years later, Yang and Mills (YM) generalized Maxwell's theory to non-Abelian fields which take their values in the Lie algebra $\fG=\su(2)$  and can thus be acted upon by $G=\SU(2)$-valued gauge transformations  \cite{YangMills, AbersLee}.
In detail, YM fields are  described by the Yang-Mills potential represented either by a $3$-vector $\bA =(A_{\mu}^a),\, a=1,2,3$ or alternatively, by  
antihermitian $\su(2)$ matrices,
$A_{\mu} =A_{\mu}^a\,\smallover{1}/{2i}\sigma_a$, where the sigmas are the Pauli matrices 
$$
\sigma_1 = {\tiny \barray{cc}0 &1 \\ 1 &0\earray}\,,
\;\;
\sigma_2 = {\tiny \barray{cc}0 &-i \\ i &0\earray}\,,
\;\;
\sigma_3 = {\tiny \barray{cc}1 &0 \\ 0 &-1\earray}\,.
$$
which satisfy  $[\sigma_a,\sigma_b]= 2i\epsilon_{abc}\sigma_c$.  
In what follows we shall use mainly the matrix-formalism.
Space-time indices will be denoted by greek letters, typically  $\mu,\nu,\dots$ etc. Latin characters $a,b\dots$  are used for the internal, isospin indices.
The Lie bracket in $\su(2)$ is $\big([A,B]\big)^a= \epsilon^a_{\;bc}A^b B^c$. 
The field strength of a Yang-Mills field is
\beq
F= \half F_{\mu\nu} dx^\mu \wedge dx^\nu \where
F_{\mu\nu} = \p_{\mu}A_{\nu}-\p_{\nu}A_{\mu}+\big[A_{\mu},A_{\mu}]\,.
\label{YMfstrength}
\eeq
The Lie algebra carries a  metric  given by the trace form, $g_{ab}A^aB^b=-2\tr(AB)$ we denote also by 
  $A\cdot B$. For $\su(2), \, g_{ab}=\delta_{ab}$.
  
The fundamental property of Yang-Mills theory is its behavior under an $\SU(2)$-valued gauge transformation \cite{YangMills}, 
\beq
A_\mu \to g^{-1} A_\mu \, g + g^{-1} \partial_\mu g\,,
\qquad
F_{\mu\nu} \to g^{-1}F_{\mu\nu} \, g\,,
\label{gaugetr}
\eeq 
where $g(x^\mu)\in\SU(2)$. 
A particle is coupled to the  electromagnetic field $A_{\mu}$  by \emph{minimal coupling}, which amounts to replacing ordinary derivatives by gauge-covariant derivatives \cite{AbersLee},
\beq
\p_{\mu} \to \p_{\mu} - i\,A_{\mu}\,,
\label{emincoupl}
\eeq
where the electric charge was scaled to one. 
In  \cite{Kerner68} Kerner argued that in a YM gauge field, this prescription should be replaced by an expression which (i) describes the properties of proton/neutron type ``particles with internal YM structure''
 (ii) couples such a particle to the non-Abelian gauge potential~: the rule \eqref{emincoupl} should be by generalized  to
\beq
\p_{\mu} \to \p_{\mu} + {\,}A_{\,\mu}
\label{YMmincoupl}
\eeq
acting on fields
in the fundamental representation.
The non-Abelian coupling constant is scaled to unity.

What is the dynamics of such an isospin-carrying particle (also called a particle with internal YM structure) ?
Kerner  answers the question by considering a non-Abelian generalization of Kaluza-Klein (KK) theory \cite{Kaluza,OKlein}.

\subsection{Abelian Kaluza-Klein theory}\label{AbKK} 
Electromagnetism and gravitation have been unified into a geometrical framework (now called fiber bundle theory) by Kaluza \cite{Kaluza} and by Klein \cite{OKlein} about 100 years ago \footnote{Our outline follows  \cite{GrossPerry}.}. 

 It is assumed that the world has \emph{four spatial} dimensions but one of the them we denote by $x^5$ has curled up to form a circle so small as to be unobservable. 
The basic assumption is that the correct vacuum is  $M^4\times \IS^1_R$, the product of four dimensional Minkowski space with coordinates $x^\mu,\, \mu=0,1,2,3$, with an internal circle of radius $R$. 

Then  general relativity in five dimensions contains
a local U(1) gauge symmetry arising from the isometry of the hidden fifth dimension.
The extra components of the metric tensor constitute the gauge fields and could be identified with the electromagnetic vector potential.

The theory is invariant under general coordinate transformations that are independent of
$x^5$. In addition to ordinary four dimensional coordinate transformations, we have a U(1) local gauge transformation
\beq
x^5\to x^5+\Lambda(x^\mu)
\label{KKgaugetr}
\eeq
 under which the $g_{\mu5}$ component transforms as a ${\rm U}(1)$  gauge field,
\begin{equation}
g_{\mu5}(x)\to g_{\mu5}(x)+\p_\mu\Lambda.
\label{(1.3)}
\end{equation}
We write the metric with indices $A = \mu, 5$ as,
\beq
ds^2 = g_{AB}dx^A dx^B \where 
g_{AB}=\left(\begin{array}{cc}
        g_{\mu\nu} + A_\mu A_\nu &A_\mu 
        \\[1pt]
        A_\nu &1
\end{array} \right)\,,
\label{5metrica}
\end{equation}
\ie,
\beq
 ds^2 =  g_{\mu\nu}dx^\mu dx^\nu+(dx^5+A_\mu dx^\mu)^2\,.
\label{5metric}
\eeq
Expressing the 5-dimensional scalar curvature $R_5$ in 4-dimensional terms, 
$R_5=R_4 +\smallover{1}/{4}F_{\mu\nu}F^{\mu\nu}$ where $R_4$ is  the $4$-dimensional curvature, the
 effective low-energy theory is described by the four-dimensional action
\begin{equation}
        S  =  -\frac{1}{16\pi G}\int d^4x\,\sqrt{-{\rm det}\,(g_{\mu\nu})}{\;}\Big(R_4 + \smallover{1}/{4}F_{\mu\nu}F^{\mu\nu}\Big),
\label{(1.7)}
\end{equation}
where 
$G={G_K}/{2\pi R}$ 
is Newton's constant. The internal radius $R$ is determined by the electric charge.
 The motion is given by a five-dimensional geodesic,
\begin{equation}
\frac{d^2x^A}{d\tau^2}+\Gamma^A_{\ BC}\,\frac{dx^B}{d\tau\; }\frac{dx^C}{d\tau\; }=0\,.
\label{(1.12half)}
\end{equation}
The KK space-time possesses a Killing vector, namely
\begin{equation}
 K^A \frac{\ \partial}{\partial x^A} =\frac{\ \partial}{\partial x^5}\,,
\label{(1.13)}
\end{equation}
which implies that 
\begin{equation}
    q = K_A\frac{dx^A}{d\tau}= 
    \frac{dx^5}{d\tau}+A_\mu\frac{dx^{\mu}}{d\tau}
\label{(1.14)}
\end{equation}
is a \emph{constant of the motion}  identified with the conserved electric charge.
The remaining equations of motion then take the form,
\begin{equation}
        \frac{d^2x^\mu}{d\tau^2}+\Gamma^\mu_{\alpha\beta}\frac{dx^\alpha}{d\tau}
        \frac{dx^\beta}{d\tau}=q\,\big(g^{\mu\alpha} F_{\alpha \nu}\big)\frac{dx^\nu}{d\tau}\,, 
\label{(1.15)}
\end{equation}
where $\Gamma^\mu_{\alpha\beta}$ is the  
Levi-Civita connection constructed from the four dimensional metric $g_{\mu\nu}$. On the right we recognize the Lorentz force. of electromagnetism.

\subsection{Non-Abelian generalization}\label{NAbKK}

Kerner, in his groundbreaking paper \cite{Kerner68}, proposed to derive the dynamics of an isospin-carrying particle in a Yang-Mills (YM) field  by generalizing the Abelian KK framework to non-Abelian gauges.
His framework was further generalized \cite{Cho75} and applied later to particle motion in a Yang-Mills field by projecting the geodesic motion to $4D$ space \cite{FeherAPH}.   His clue [eqn. \#(12) of \cite{Kerner68}] is to replace  the internal circle U(1) in the 5th dimension by the non-Abelian gauge group, $\SU(2)$ and the gauge potential in \eqref{5metrica} by its non-Abelian counterpart.
The key new ingredient w.r. t. electromagnetism is the \emph{isospin}, represented by an $\su(2)$ matrix,
\beq 
Q = Q^a \,\frac{1}{2i}\sigma_a \in \su(2)
\,,
\label{Qvector}
\eeq
which couples the particle to the YM field introduced  in sec.\ref{YMSec},
  $A^a_{\,\mu}$ and  $F^a_{\;\alpha\beta}$, respectively.
The covariant derivative is 
\beq
D_{\mu}Q = \p_{\mu}Q + [A_\mu, Q]\,.
\label{DQaction}
\eeq
The $\su(2)$-valued YM potential  is implemented on the isospin $Q\in \su(2)$ by commutation.
 
 In a judicious coordinate system chosen by Kerner \cite{Kerner68},
the equations of motion for a test particle in the combined gravitational and gauge fields simplify to his eqn. \# (34),
\beq
\bigbox{
\frac{d^2x^{\mu}}{ds^2}+\Gamma^{\mu}_{\alpha\beta}
\frac{dx^{\alpha}}{ds}\,\frac{dx^{\beta}}{ds}
= \big(g_{ab}Q^b\big)\,\big(g^{\mu\alpha}F^a_{\;\alpha\beta}\big)\frac{dx^{\beta}}{ds} \,.
}
\label{Kernereq}
\eeq
 
Generalizing the gauge group from U(1) of electromagnetism to the Yang-Mills gauge group $\SU(2)$ has a price to pay, though: unlike the electric charge in the electromagnetic theory which is a conserved scalar, the isospin has indeed its own dynamics~: it is \emph{not} a constant but a vector
which (as Kerner  puts it) ``{\sl rotates, depending on the external field}''.
 
The equations for the  motion of the isospin,
\beq
  \dot{Q} =
[{Q},A_\nu \dot{x}^\nu]\,, 
\label{isospineq}
\eeq
[where
the ``dot'' is $ d/{ds}$]
were spelt out two years later by Wong \cite{Wong70}. In a geometric language,  the isospin is parallel transported along the space-time trajectory, $x(t)=(x^\mu)$. Written in terms of the covariant derivative \eqref{DQaction},
\beq
\medbox{
D_s{Q} \;\equiv\;  \dot{Q} 
 + 
\left[A_\nu \dot{x}^\nu,{Q}\right]
=0\, 
}
\label{covisoseq}
\eeq
this equations says that the isospin is \emph{covariantly} (but not ordinarily) conserved. 
Eqn. \eqref{isospineq} is consistent with Kerner's words, though, and also with what  Yang and Mills say in their \cite{YangMills}, where they mention ``isospin rotation''.
    
One can wonder why did Kerner \emph{not} spelt out  the equations of motion for the isospin explicitly. A real answer can be given only by him, however one can try to guess what he might have had in his mind.  
One good reason might well have been that  considering the isospin as a non-constant non-Abelian analog of the constant electric charge could have appeared too radical and even shocking, and be therefore discarded 
\footnote{ Duval's note \cite{DuvalCRAS} was rejected from   Comptes Rendues de l'Acad\'emie des Sciences \emph{without refereeing}.}.

There might exist also other, subtle reasons related to the \emph{gauge invariance} and the consequent problems of physical interpretation \cite{Arodz82,HRawnsley,HRcolor}.
Another one could come from the experimental side.  
 
Wong's approach \cite{Wong70} is radically different from that of Kerner~: instead of generalizing the classical dynamics of a charged particle moving in a curved space, he ``dequantizes'' the Dirac equation. Balachandran et al. \cite{Balachandran76,Balachandran77}, studied particles with internal structure which were then  recast in a symplectic framework by Sternberg
\cite{Sternberg77,Sternberg78,Sternberg80}, by Weinstein \cite{Weinstein78}, and by Montgomery \cite{Montgomery}.
Duval \cite{DuvalCRAS,DuvalAix79,DH82} extended Souriau's approach \cite{SSD} to  particles with spin 
\cite{DuvalAix79} 
 --- hitting yet another shocking idea: physicists, referring to Landau-Lifshitz, were  firmly convinced that \emph{classical spin just does not exist} and rejected Souriau's ideas \cite{SSD}
 rooted in the representation theory. 
 
Gauge fields with spontaneous symmetry breaking admit finite-energy static solutions with  magnetic charge referred to as \emph{non-Abelian monopoles}  \cite{GoddardOlive}.  
For an isospin-carrying particle in the field of a selfdual monopole  \cite{Feher:1984ik,Feher:1984xc,Feher86
 } Feh\'er found, moreover, that outside the monopole core, where the $\SU(2)$ symmetry is spontaneously broken to ${\rm U}(1)$, the dynamics of a particle with isospin reduces to that of an electrically charged particle in the field of a Dirac monopole, combined with specific scalar potentials, familiar from the Abelian theory \cite{MIC,Zwanziger68}.

\subsection{Fibre bundles and a symplectic framework}\label{symplecticSec}

Trautman  \cite{Trautman70}, and Cho \cite{Cho75} reformulated the non-Abelian KK theory in terms of fibre bundles  \cite{Kobayashi}~: for gauge group $G$, the  field is described by a Lie algebra-valued connection form  $\alpha$ on a principal bundle $\IP$ with structure group $G$  over space-time, $M$. The YM potential $A$ in sec.\ref{Kernerfw} is the pull-back to $M$ of the connection $1$-form by a section  $M \to \IP$ of the bundle. A gauge transformation amounts to changing the section and results  in  \eqref{gaugetr}. Choosing a section yields a local trivialisation $\IP = M \times G$ and the YM connection form is written as,
\beq
\alpha = A_\mu dx^\mu + g^{-1}dg\,.
\label{YMconn}
\eeq
Recall that  the Maurer-Cartan form $g^{-1}dg$ takes its values in Lie algebra $\fG$ of $G$.
Using fiber bundles for gauge theory was advocated by T. T. Wu and C.N. Yang \cite{WuYang75, WY76,CNY79} in the monopole context \footnote{
Souriau has discussed the fiber bundle description of a monopole in ``Prequantization'' chapter of his never completed and thus unpublished revision of his book \cite{SSD}.}; see also \cite{BalaMarmo79,HPAAix79}.
\goodbreak

A comprehensive KK unification of \NA gauge fields with gravity in  principal fibre bundle terms was put forward by Cho in \cite{Cho75}, who 
 derived a  unified Einstein-Hilbert action in (4+n)-dimensions  both in the basis used by Kerner and also in a horizontal-lift basis which diagonalizes the KK metric and  generalizes  \eqref{(1.7)}. 

\begin{figure}[h]
\includegraphics[scale=.5]{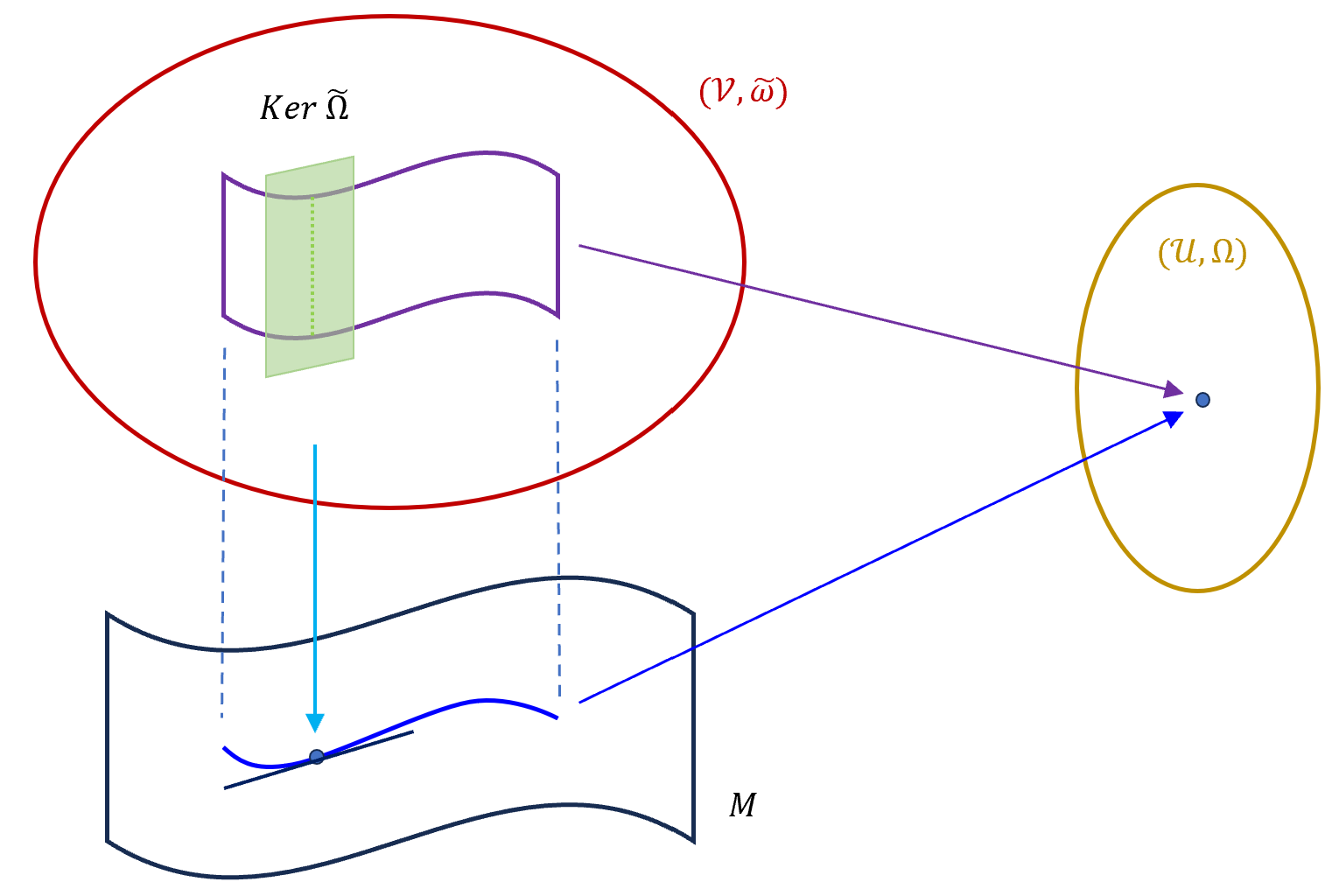}
\vskip-3mm
\caption{\textit{\small Souriau's framework: the {\bf worldline} in ${\bf M}$ is the projection of a \cyan{\bf characteristic sheet} of the 2-form \magenta{$ 
\widetilde{\bf \Omega} = {\bf d\tilde{\bomega}}$} on the \magenta{\bf evolution space}, \magenta{$\cV$}. 
Factoring out the characteristic foliation tangent to  
\dgreen{$\bf{\ker}\,\tilde{\bOmega}$}, $\cV$ projects to the \orange{\bf space of motions}, 
\orange{$\cU$}, to which the 2-form $d\tilde{\omega}$ projects as a symplectic form 
\orange{${\bf \Omega}$} and correspond to  
the worldlines in $M$.
}}
\label{SympFig}
\end{figure}

Duval et al \cite{DuvalAix79, DH82}  proposed an alternative, symplectic version ``\`a la Souriau'' \cite{SSD}, reminiscent of but different from the Kaluza-Klein   approach. Both theories use a higher-dimensional, fiber bundle extension of the conventional space-time structure. Below we summarize the main features of the Souriau framework~:
\begin{enumerate}
\item 
The system is described by a fiber bundle $\cV$ over space-time $M$ called an  \emph{evolution space} -- Souriau's ``espace d'\'evolution'';

\item The dynamics is discussed  in terms of differential forms. The main tool is a $1$-form $\tilde{\omega}$ on $\cV$ whose exterior derivative $\tOmega= d\tomega$ is, in Souriau's language, ``presymplectic'', \ie, a closed 2-form which  has constant rank,  dim $\kernel\tOmega = \const$. Then the  motions are the projections onto $M$ of the  integral submanifolds of the characteristic foliation of $\kernel\tOmega$. Factoring out $\kernel\tOmega$ yields $\cU$, the \emph{space of motions} (an abstract substitute for phase space --- Souriau's ``espace des mouvements'' \cite{SSD}). The presymplectic form $\tOmega$ projects onto $\cU$ as a symplectic form \ie, one which is closed and has no kernel, as illustrated in FIG. \ref{SympFig}.

\item
A group $S$ is a symmetry for a system if it acts on the space of motions  $\cU$ by preserving the symplectic structure.
 
\item
A system is \emph{elementary} with respect to a symmetry group $S$ if the action of the latter on $\cU$ is transitive. Souriau's orbit construction  \cite{SSD} applies to an arbitrary symmetry group: the space of motions of an elementary system is, conversely, a (co)adjoint orbit $\cO=\big\{g^{-1} Q_0g \,\big|\, g\in S\,  \big\}$ 
of a basepoint $Q_0$ chosen  in the Lie algebra $\in \fS$ of the symmetry group. 
$\cO$  is endowed with its canonical symplectic form,
\beq
\tOmega = d\tomega, \qquad
\tomega = Q_0\cdot(g^{-1}dg)\,.
\label{canconnsymp}
\eeq
In particular, applying the general the construction to the gauge group, $G$, endows the orbit in dual of the Lie algebra $\fG$  with its canonical symplectic form.

\item
The symmetry group $S$ w.r.t. which the system is elementary can be viewed itself as evolution space, $\cV=S$ \cite{Kunzle72}; $S$ is a principal fiber bundle over its (co)adjoint orbit $\cO$.
\end{enumerate}

Now we spell out a simplified form of the Souriau-Duval framework in flat space. For further details the reader is advised to consult \cite{DuvalAix79,DH82}.
\goodbreak

\medskip
$\bullet$ \underline{A massive free relativistic particle}. The Poincar\'e group ($P$) is a fiber bundle over Minkowski spacetime $M$ with the Lorentz group as structure group  \cite{Kunzle72,DuvalAix79}.
We represent the Poincar\'e group by $5\times5$ matrices
$
{\barray{ll} L & x \\ 0 &1\earray}
$
where the $4\times 4$ matrix $L$ belongs to the Lorentz subgroup  and
$x = (x^\mu) \in M$. Then  
\beq
\text{Poincar\'e} / \text{Lorentz} = M = \text{Minkowski}\,.
\label{PLMink}
\eeq
Moreover, we choose the basepoint $Q_m$ in the Poincar\'e Lie algebra, 
\beq
Q_m=\barray{cc} 0 & x_m \\ 0 &0\earray
\with x_m = m{\tiny \barray{l} 0 \\ \vdots \\1\earray}\in M
\label{Pbasepoint} 
\eeq
 where $m=\const$ interpreted as rest mass. Writing the Maurer-Cartan form as
\beq
g^{-1} dg =\barray{cc}
 L^{-1}dL & L^{-1}dx \\ 0 &0\earray
\label{PMaurercartn} 
\eeq
 we get, on the Poincar\'e orbit $\cO_m$ of $Q_m$,
\beq
\tomega_m = m I_\mu dx^\mu \Rarrow \tOmega_m = 
m dI_\mu \wedge dx^\mu\,
\label{Pforms}
\eeq
where $I_\mu$, a component of the Lorenz matrix, is future pointing and belongs to the unit tangent bundle of $M$ \cite{Kunzle72,DuvalAix79,DH82}. 
Then the characteristic foliation projects, in a suitable parametrisation, to $M$ onto a curve, which is a solution of 
\beq
\dot{x}^\mu = I^\mu,
\qquad
\dot{I}^\mu = 0\,.
\label{freeqmot}
\eeq
 Eqn. \eqref{freeqmot} describes the geodesic motion in Minkowski space --- \ie, the motion of a free relativistic particle with no spin \footnote{Spinning particles are obtained by modify the basepoint $Q_0$ in \eqref{Pbasepoint}, cf. eqn. \#(3.9) in \cite{DH82}.}.

$\bullet$ The free theory based on the Poincar\'e group $P$ is readily extended to a (still free) relativistic particle with internal structure: 
 enlarging the evolution space and 1-form, $P$ and $\tomega_m$, respectively, to
\beq
\IP = P \times G
\aand
\omega = \tomega_m + Q_0\cdot g^{-1}dg\, ,
\label{Womega}
\eeq
 where $g$ takes its values in the gauge group $G$ and the basepoint is $Q_0 \in\fG$ (the Poincar\'e part being understood).
 
 The kernel of $\tilde{\Omega}$ 
in \eqref{canconnsymp} implies the free equation
 \eqref{freeqmot}, supplemented by that for the isospin, 
 \eqref{covisoseq}, whose properties will be further studied in sec. \ref{IsochargeSec}. 
 In geometric language, the isospin belongs to the associated bundle $\IP\times_G\cO_0$,
where $\cO_0$ is the (co)adjoint orbit of $Q_0$ in $\fG$. In local coordinates,
$ \IP\times_G\cO_0 \simeq  M\times \cO_0\,$ \cite{DH82}.
The space of motions is
$\cO_m\times\cO_0$
endowed with the projection of $\tilde{\Omega}$ in \eqref{canconnsymp}.

\medskip
For $G=$ U$(1)$ the free charged particle is recovered, with $Q$ identified with the constant electric
charge.

$\bullet$ minimal coupling to a Yang-Mills field amounts, in bundle language, to generalize \eqref{Womega} on $\IP$ by,
\beq
\omega = \tomega_m + Q_0\cdot \alpha\,, 
\label{alphaomega}
\eeq
with $Q_0 \in \fG$,
which, in view of \eqref{YMconn}, is indeed the geometric form of \eqref{YMmincoupl}. In local coordinates,
\beq
\omega=\big(\p_{\mu}+Q_0\cdot A_\mu\big)dx^\mu 
+ Q_0\cdot g^{-1}dg\,.
\label{localconnform}
\eeq

The 2-form $\Omega=d\omega$ is, by the Cartan structure equations \cite{Kobayashi}, p.78,
\beq
\Omega =
\Omega_{0} + Q_0 \cdot d\alpha = \Omega_{0} +  Q_0 \cdot \big(D\alpha - [\alpha,\alpha]\big)\,,
\label{coupled2form} 
\eeq
where the last term involves, in addition to the Lie bracket, also the wedge product of the differential forms. Its
 kernel projects to the Kerner-Wong equations
\eqref{Kernereq}-\eqref{isospineq}  \cite{DH82}. 

\section{Physical meaning of isospin dynamics}\label{IsochargeSec} 

Limiting our investigations to flat Minkowski space,  
the Kerner equations \eqref{Kernereq} simplify to,
\beq
\ddot{x}_\mu =  Q^a{\,}F_{\mu\nu}^a\,\dot{x}^\nu\,,
\label{flatKerner}
\eeq
supplemented by the isospin equation \eqref{isospineq}
\footnote{
The equations \eqref{flatKerner}-\eqref{isospineq} were also studied  by refining the field-theoretical arguments of Wong  \cite{Arodz82}. The classical isospin is the expectation value of the non-Abelian field, $Q^a=\half\int\!\psi^{\dagger}\sigma_a\psi$\,.}.

To what extent is the isospin vector, ${Q}$, an analog of the \emph{constant}  electric charge ?
We argue that ${Q}=\const$  would be inconsistent with gauge invariance: if we had $\dot{Q}=0$, 
the rhs of \eqref{isospineq} would change, under a gauge transformation, as,
$$
0= [{Q},A_{\mu} \dot{x}^{\mu}] \to 
[{Q},(g^{-1}A_{\mu}g) \dot{x}^{\mu}]
+
[{Q}, (g^{-1}\p_{\mu}g) \dot{x}^{\mu}]\,,
$$
and there is no reason for the rhs to vanish. 
 The situation improves, though, if the gauge transformation  is non-trivially implemented on the isospin \footnote{
 The covariant  transformation rule \eqref{gaugeonIso}  is consistent with the geometric status of the isospin viewed as a section of the associated bundle $\IP\times_G\cO$ \cite{DH82,HP-Kollar}.},
\beq
{Q} \to  g^{-1} {Q} \, g\,.
\label{gaugeonIso}
\eeq 
Then the rhs of \eqref{isospineq} would transform as
$$
[{Q},A_{\mu} \dot{x}^{\mu}] \to 
g^{-1}\Big\{[{Q},A_{\mu} \dot{x}^{\mu}]+ \left({Q}\p_{\mu}g\,g^{-1}-\p_{\mu}g\,g^{-1}{Q}\right)\dot{x}^{\mu}\Big\}\,g\, .
$$
The first term in the curly bracket would be perfect but the 2nd one would vanish only for  $g=\const$.
However the terms coming from $d(g^{-1} {Q} \, g)/ds$   cancel the unwanted terms, leaving us with the desired covariant transformation law cf. \eqref{gaugeonIso},
\beq
D_{s}{Q}  \to g^{-1} D_{s}{Q}\, g\,.
\label{covisoeq}
\eeq

Further insight into isospin dynamics can be gained by assuming, for simplicity, that the curvature of the connection form (in physical terms, the Yang-Mills field) is zero, $F = D\alpha=0$ which is a gauge-independent statement by \eqref{gaugetr}, and the space-time motion is free. 
Do we have also
$\dot{Q} = 0\; ?$
The answer is: \emph{yes and no}. Let us explain. In topologically trivial situations \footnote{The topologically non-trivial case is studied in \cite{HPNABA,HP-Kollar,HP-EPL}.},
 $F = D\alpha=0$ implies that one can find a gauge where $A_{\mu} = 0$ and then $\dot{Q} = 0$ follows obviously from the isospin equation \eqref{isospineq}.
This is a \emph{gauge-dependent statement}, though~: We are  allowed to apply a gauge transformation by an arbitrary $G = \SU(2)$-valued function $g(x^\mu)$ which   changes $A_{\mu} = 0$ to a pure gauge $A_{\mu} = g^{-1}\p_{\mu}g$ -- \emph{but it rotates also the isospin}, \eqref{gaugeonIso}. 
$
d{\big(g^{-1} {Q} \, g\big)/ds}\neq0
$
in general; the gauge-covariant statement is that the isospin is {covariantly} conserved, \eqref{covisoseq}. 

What is then the physical  meaning of the isospin vector~? First we note that 
\beq
\vert Q\vert^2 = Q^aQ^a
\label{Qsquare}
\eeq
 is gauge invariant, and  deriving it implies,  using \eqref{covisoeq}, 
that the length $\vert {Q}\vert$ is conserved, 
\beq
\frac{d|{Q}|}{ds} = 0 \Rarrow  |{Q}|=\const.
\label{Ilength}
\eeq
The isospin is thus constrained to lie on an adjoint orbit of the gauge group $G$ in its Lie algebra $\fG$ -- in our case, to a {sphere},
$
{Q}\in \cO = \Big\{g^{-1} Q_0 g \,\Big|\, g \in \SU(2)\Big\} \approx \IS^2\,.
$
It is  this fact that is behind the Souriau-type construction of isospin-extended models \cite{DuvalAix79,DH82,HP-Kollar}.

Which  components of ${Q}$ do have a gauge invariant physical meaning ? -- the question leads to the so-called ``color problem'' \cite{NelsonMano83,MarmoBala82,NelsonColeman84}.
The point is the subtle difference between gauge transformations and internal symmetries  \cite{HRawnsley,HRcolor}. 

In physical terms: can we implement an element of the gauge group on the physical fields~? And if we can, will it be a symmetry in the usual sense \cite{ForgacsManton} ? In bundle language, a gauge transformation acts on the fibers \emph{from the right} \cite{Kobayashi}, --- while a symmetry should act \emph{from the left} \cite{HRawnsley,HRcolor}. Can we transfer the right-action to a left action ? 
In geometric terms, ``implementable'' means that the $G=\SU(2)$ bundle ${\IP}$ should be reducible, and ``symmetry'' requires that the connection form $\alpha$ in \eqref{YMconn} which represents the
YM potential should also be reducible to the reduced bundle.

When the underlying topology is non-trivial (as non-Abelian monopoles \cite{GoddardOlive}), there can be an obstruction~: \textit{global color can not  be defined''}, as it is put in refs. \cite{NelsonMano83,MarmoBala82,NelsonColeman84}. 
Another  physical instance is provided by the Non-Abelian Aharonov-Bohm effect \cite{WuYang75}, for which there is no obstruction but there is an ambiguity of how it should be implemented \cite{EZH-NABA}. 

\section{Conservation laws with Isospin}\label{ConsQuant}

\subsection{van Holten's covariant framework}\label{vHalgo}

The Hamiltonian  of a point particle of unit mass carrying isospin $\vec{Q}=(Q^a)$ which moves in a static YM field is,
\begin{equation}
H=\frac{1}{2}\left(\vp-\vA^{a}Q^{a}\right)^{2}\,,
\label{isomagham}
\end{equation}%
where we scaled the coupling constant again equal to one.
Defining the covariant Poisson bracket  as \cite{vHolten},
\beq
\big\{f,g\big\} = \,D_jf\,\frac{\p g}{\p \pi_j}-\frac{\p f}{\p \pi_j}\,D_jg 
+\;Q^aF_{jk}^a\frac{\p f}{\p \pi_j}\,\frac{\p g}{\p \pi_k}
-f^{abc}\frac{\p f}{\p Q^a}\frac{\p g}{\p Q^b}\,Q^c,
\label{PBracket}
\eeq
where the $f^{abc}$ are the structure constants of the Lie algebra, and the
covariant phase-space derivative is,%
\begin{equation}
D_{i}f=\p_{i}f-f^{abc}Q^{a}A_{i}^{b}\frac{\partial f}{\partial Q^{c}}\,.
\end{equation}%
The nonzero Poisson brackets are, %
\begin{equation}
\left\{x_{i,}p_{j}\right\} =\delta _{ij},\text{ \ \ }\left\{
Q^{a},Q^{b}\right\} =-f^{abc}Q^{c}\,.
\end{equation}%
Then the Hamilton equations with $\vpi=\dot{\br}$,
\beq
\dot{x}_i=\big\{x_i,H\big\}, 
\qquad 
\dot{\pi}_i=\big\{\pi_i,H\big\}, 
\qquad
 \dot{Q}^a=\big\{Q^a,H\big\}\,
\eeq 
allow us to recover the flat-space Kerner-Wong equations, 
\besub   
\begin{align}
\ddot{x}^{i}=& \; Q^{a}F_{ij}^{a}\dot{x}^{j} 
=Q^{a}\varepsilon^{ijk}\dot{x}^{j}B_{a}^{k}\,,%
\label{Kernerbis}
\\[3pt]
\dot{Q}^{a}=& \;f^{abc}\dot{x}^{i}A_{b}^{i}Q_{c}\,,
\label{isobis}
\end{align}
\label{KernerWongeq}
\esub
 equivalent to \eqref{flatKerner} and \eqref{isospineq}.

Following van Holten \cite{vHolten,vHolten2,vHolten3,HP-NGOME},
constants of the motion can  be sought for by expanding into powers of the covariant momentum, 
\beq
q=C(\vx)+C_i(\vx)\,\pi_i+\frac{1}{2!}C_{ij}(\vx)\,\pi_i\pi_j+\dots
\label{consquant}
\eeq

Skipping the Abelian case, we move directly to the non-Abelian one. Requiring $q$ to Poisson-commute with the Hamiltonian then yields a series of constraints,  eqn. \# (70) in \cite{vHolten}. 
\begin{equation}
\begin{array}{ll}
D_iC = Q^aF_{ij}^a C_j\,,
\\[3pt]
D_iC_k+D_jC_i=Q^a\big(F_{ik}^a C_{kj}+F_{jk}^aC_{ki}\big)\,,
\\[3pt]
D_iC_{jk}+D_jC_{ki}+D_kC_{ij}=Q^a\big(
F_{il}^aC_{ljk}+F_{jl}^aC_{lki}+F_{kl}^aC_{lij}\big)\,,
\\
\qquad\qquad\vdots\qquad\qquad\qquad\qquad\qquad\qquad\qquad\vdots&\qquad
\end{array}
\label{vHconstraints}
\end{equation}

The expansion \eqref{vHconstraints} can be  truncated at a finite order when the covariant Killing equation is satisfied at some order $n$. When we have a Killing tensor,
$ 
D_{(i_1}C_{i_2\dots i_n)}=0\,,
$ 
then we can set 
\beq
C_{i_1\dots i_{p}}=0
\eeq
for all $p \geq n$, 
and find a constant of the motion of the polynomial form,
\beq
q= \sum_{k=0}^{p-1}\frac{1}{k!}C_{i_1\dots i_k}\pi_{i_1}
\dots \pi_{i_k}\,
\label{truncseries}
\eeq
\cite{vHolten}.
Referring to the literature for details \cite{vHolten,vHolten2,vHolten3,HP-NGOME} we mention that in
the Abelian theory  $\vec{Q}$ is just a constant identified with the electric charge. 

\smallskip
The van Holten algorithm can be generalized by adding
a static scalar potential which may depend also on the isospin. The  Hamiltonian \eqref{isomagham} then becomes\beq
H=\frac{1}{2}\pi _{i}^{2}+V\left(x^{i},Q^{a}\right)
\label{isoVHam}
\eeq
with equations of motion,
\besub
\begin{align}
\ddot{x}^{i}= \, &\, Q^{a}F_{ij}^{a}\dot{x}^{j}-D_{i}V\,,
\label{pieq}
\\[2pt]
\dot{Q}^{a}= \,&\, f^{abc}\dot{x}^{i}A_{b}^{i}Q_{c}+f^{abc}Q^{b}\frac{\partial V}{%
\partial Q^{c}}\,.
\label{isoeqn}
\end{align}
\label{KWeq}
\esub
Comparison with  \eqref{KernerWongeq} then shows that \eqref{pieq}
 picks up a covariant force term. Note also that when $V$ does depend on $\vI$ the isospin is not more parellel transported. 
 
Generalizing \eqref{consquant} to isospin-dependent coefficients,
\begin{equation}
q(\vx,\vI)=C(\vx,\vI)+C_i(\vx,\vI)\,\pi_i+\frac{1}{2!}C_{ij}(\vx,\vI)\,\pi_i\pi_j+\dots
\label{consquantNA}
\end{equation}
the constraints
\eqref{vHconstraints} are also generalised \cite{HP-NGOME},
\begin{equation}
\begin{array}{lll}
\;C_iD_iV+\displaystyle{f^{abc}Q^a\frac{\p C}{\partial Q^b}\frac{\p V}{\p Q^c}}=0\,,
\\[12pt]
D_iC=Q^aF^a_{ij}\,C_j+C_{ij}D_jV+\displaystyle{f^{abc}Q^a\frac{\p C_i}{\partial Q^b}\frac{\p V}{\p Q^c}}\,,
\\[10pt]
D_iC_j+D_jC_i=Q^a(F^a_{ik}C_{kj}+F^a_{jk}C_{ki})+\;C_{ijk}D_kV
+ \displaystyle{f^{abc}Q^a\frac{\p C_{ij}}{\partial Q^b}\frac{\p V}{\p Q^c}}\,,
\\
\quad\qquad\vdots\qquad\qquad\quad\qquad\qquad\qquad\qquad\vdots&\qquad
\end{array}
\label{constraints}
\end{equation}
New, gradient-in-$V$ terms thus arise even when the potential does not depend on the isospin, $V=V(\vx)$. These terms play a r\^ole for self-dual Wu-Yang monopoles \cite{WuYang69}, and for diatoms \cite{MSW}, as it will be seen in subsections \ref{WYmonopSec} and \ref{DiatSec}, respectively.

\benu
\item
When $C_i(\vx)$ is a Killing vector then we have
$p=2$ and the expansion can be reduced to a linear expression,
\beq
q=C(\vx)+C_i(\vx)\,\pi_i\,,
\eeq
 allowing us to recover the conserved momentum and angular momentum \cite{vHolten}. Focusing our attention at the latter, we choose a unit vector $\vn$; then 
\beq
\vC=\vn\times\vx\, 
\label{rotKilling}
\eeq
is a Killing vector  for rotations around $\vn$ and thus  generates the conserved angular momentum, $\vJ$.
 van Holten's recipe can be applied also to a Dirac monopole of charge $q=eg$, recovering the angular momentum vector
\beq 
\vJ=\vx\times\vpi-q \,\hx\, ,
\label{Diracangmom}
\eeq
which includes the  celebrated radial ``spin from isospin'' term \cite{Hasi,HP-NGOME}.
 
\begin{figure}[ht]
\includegraphics[scale=.66]{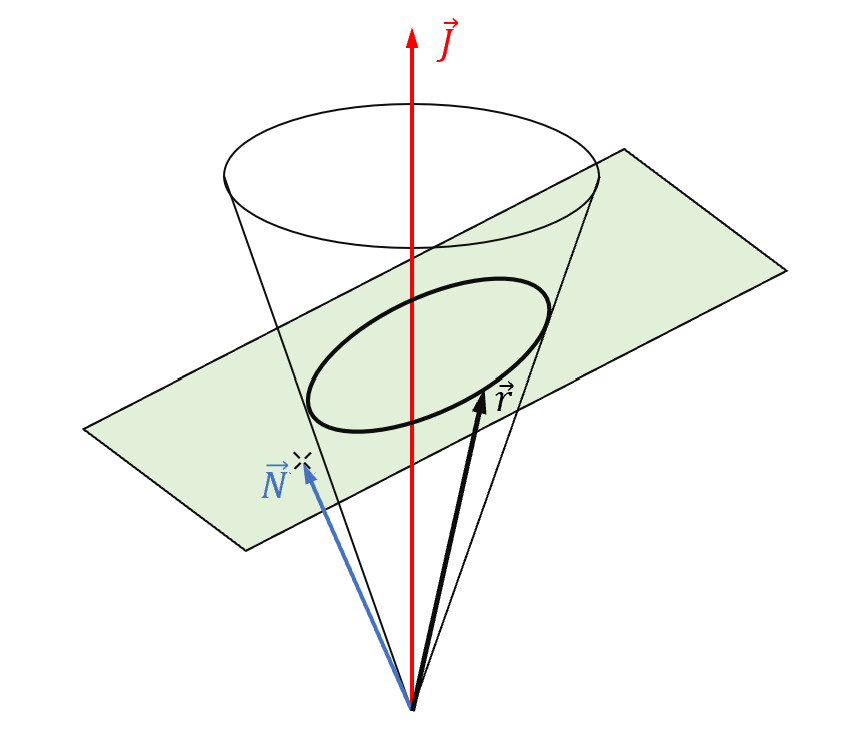}
\vskip-5mm
\caption{\textit{\small The conservation of the monopole angular momentum $\vJ$ 
 implies that a 
particle moves on a cone, whose axis is $\vJ$. The ${\rm O}(4)/{\rm O}(3,1)$ dynamical symmetry generated by the angular momentum and the Runge-Lenz vector $\vK$ implies in turn that the trajectory lies in the plane perpendicular to $\vN = \vK + (\alpha/q) \vJ$ and is therefore a conic section.
}}
\label{conicsecmot}
\end{figure}

\item 
Similarly, choosing again a unit vector $\vn$,
\begin{equation}
C_{ij}=2\delta_{ij}\, \vn\cdot\vx-(n_ix_j+n_jx_i)
\label{RLKilling}
\end{equation}
is a Killing tensor of order $2$ which generates the well-known Runge-Lenz vector of planetary motion, \cite{vHolten,vHolten2,vHolten3,HP-NGOME},
\beq
\vK=
\vpi\times{\vJ}+\alpha\,{\hx}\,.
\label{RungeLenzK}
\eeq
 More generally, the framework applies also to the so-called ``MIC-Zwanziger'' system \cite{MIC,Zwanziger68}, which combines a Dirac monopole of charge $q$ with an arbitrary $r^{-1}$ Newtonian and a fine-tuned inverse-square  potential,
\beq
V(r)=\frac{q^2}{2r^2}+\frac{\alpha}{r}\,.
\label{MICZpot}
\eeq
The combined system generalizes the well-known dynamical O(4)/O(3,1) symmetry of planetary motion spanned by the angular momentum 
and the Runge-Lenz vector, $\vJ$ in \eqref{Diracangmom} and $\vK$, respectively \cite{MIC,Zwanziger68}.
The relations
\beq
\vJ\cdot\hx= -q
\aand
\big[\vK +\smallover{\alpha}/{q} \vJ\big]\cdot\vx = \vJ^2 - q^2
\eeq
then imply that the motion is a conic section,  as depicted in FIG.\ref{conicsecmot}. 

Spin can also be considered \cite{Feher:1988th}.

We mention that the MIC-Zwanziger system is essentially equivalent to the one which describes long-range monopole scattering \cite{GibbonsManton86}
alias Kaluza-Klein monopole \cite{GrossPerry,Sorkin} See also \cite{FH86,Feher86,Feher87,Cordani88,CFH90,Feher:2009wwp,MantonSutcliffe}. 
The dynamical symmetry will  be further analysed for a self-dual Wu-Yang monopole \cite{HWY} in the next subsection.
 
\item  
  The van Holten algorithm applies also to quantum dots, H\'enon-Heiles and Holt systems, with Killing tensors whose rank  ranges from one to six are studied in \cite{vHolten2,vHolten3}.
\eenu

\subsection{Motion in the Wu-Yang monopole field}\label{WYmonopSec}

The Wu-Yang monopole \cite{WuYang69} is given by 
the non-Abelian gauge potential with a ``hedgehog'' magnetic field,
\begin{eqnarray}\displaystyle{
A_i^{a}=\epsilon_{iak}\frac{x_k}{r^2}}\ ,
\qquad
F_{ij}^a =\epsilon_{ijk}\,\frac{x_k\;x_a}{r^4}\ .
\label{WYmonopA}
\end{eqnarray}
The terminology is justified by presenting the field strength as
\beq
B_k^a = \half\epsilon_{ijk}F_{ij}^a=\frac{x^kx^a}{r^4}\,.
\label{magYM}
\eeq
The projection of the Wu-Yang magnetic field onto the ``hedgehog'' direction is thus
\beq
B_k^a \cdot \hat x^a = \dfrac {x^k} {r^3}\,,
\eeq
which shows that the Wu-Yang magnetic field is that of a Dirac monopole of unit charge, embedded into isospace.  
The remarkable feature of this expression is that the external and internal coordinates are correlated.

Let us consider an isospin-carrying particle moving in a Wu-Yang monopole field augmented with a rotationally invariant  scalar potential
$V(r)$, and inquire about conserved quantities.

$\bullet$ A most important observation says that, for an \emph{arbitrary} radial potential $V(r)$, 
we can choose $C=\vec Q \cdot \hx$ which is covariantly constant, 
\beq
D_i C =0.
\eeq
and the \eqref{constraints} are satisfied  with $C_i =C_{ij}=...=0$ . 
 The van Holten algorithm
then applies, proving that the projection of the isospin onto the radial direction,
\begin{eqnarray}
q= C = \vI\cdot{\hx} 
\label{echarge}
\end{eqnarray}
is a constant of the motion  \cite{vHolten}. 

The Wu-Yang Ansatz \eqref{WYmonopA} played an important r\^ole  in later developments as it prefigured the finite-energy non-Abelien monopoles \cite{tHooft,Polyakov,GoddardOlive}.
 The  ``hedgehog" is the large-r behavior of the Higgs field, and \eqref{echarge} is identified with the electric charge outside the monopole core. See e.g. \cite{GoddardOlive} or \cite{MantonSutcliffe} for comprehensive reviews.

$\bullet$ Applied to the Killing vector (\ref{rotKilling}), we get the \emph{conserved angular momentum} \cite{vHolten},
\begin{eqnarray}
\vJ=\vx\times\vpi-q\,{\hx}\;,
\label{WYangmom}
\end{eqnarray}
which looks formally identical to the Abelian expression \eqref{Diracangmom}. Remember however that $q$ here is \emph{not} a universal constant  but the (conserved) \emph{projection of the isospin} onto the ``hedgehog'' direction $\hx$, which mixes internal and external coordinates. Thus we have the familiar radial term -- but now in the non-Abelian context.

$\bullet$  We now inquire about quantities which are quadratic in the momentum. 
Inserting (\ref{RLKilling}) into (\ref{constraints}), from the 2nd-order equation we find, 
\begin{eqnarray}
\vC=\vec{n}\times(q\,{\hx})\;. 
\label{WYvC}
\end{eqnarray}
 
For which potentials do we get a quadratic conserved quantity ? Referring to \cite{vHolten,HP-NGOME} for  details, we just record the answer:
\begin{equation}
C=\alpha\;\vec{n}\cdot{\hx}
\aand
V(r)=\frac{q^2}{2r^2}+\frac{\alpha}{r}+\beta\,,
\label{goodpot}
\end{equation}
where $\alpha$ and $\beta$ are arbitrary constants. 
The coefficient of the $r^{-2}$ term is correlated with the conserved charge $q$  \eqref{echarge} as \eqref{MICZpot} for MIC-Zwanziger \cite{MIC, Zwanziger68,Feher87}.
Collecting our results,
\begin{eqnarray}
\vK=
\vpi\times{\vJ}+\alpha\,{\hx}\,,
\label{WYRL}
\end{eqnarray}
is a conserved Runge-Lenz vector for an isospin-carrying particle in the  
Wu-Yang monopole field combined with the fine-tuned  potential $V(r)$ in (\ref{goodpot}). 

The  conserved quantities $\vJ$ and $\vK$ span an ${\rm O}(4)/{\rm O}(3,1)$ dynamical symmetry which allow us to describe the large-r motion both classically and quantum mechanically \cite{Feher86,Feher:1984xc}. The  trajectories are again conic sections as for MIC-Zwanziger in FIG.\ref{conicsecmot}.

This generalizes the Abelian result to an isospin-carrying particle outside the core of a self-dual non-Abelian monopole \cite{HWY}. 
This ``coincidence'' is explained as follows~: for large $r$,  
the gauge field of a self-dual non-Abelian monopole of  
charge $m$ \cite{GoddardOlive} is of the radially symmetric Wu-Yang form, eqn. (\ref{WYmonopA}), completed with a ``hedgehog'' Higgs field, 
\begin{equation}
\Phi^a=\Big(1-\displaystyle\frac{m}{r}\Big)\, \displaystyle\frac{x^a}{r}\,, 
\label{hedghogWY}
\end{equation}
whose direction is, precisely,
\beq
\widehat{\Phi}=\frac{\Phi}{|\Phi|} = \hx \, .
\label{Phir}
\eeq
The projection of the isospin onto $\widehat{\Phi}$, $q$ in \eqref{echarge}, is thus conserved, and outside the core the motion is that of an electric charge  in the MIC-Zwanziger  field
 \cite{MIC,Zwanziger68,Feher:1984xc,Feher86,Feher87}. The isospin-dependent dynamical symmetry is analyzed in \cite{HWY}.

\subsection{Diatomic molecules}\label{DiatSec}

In Ref. \cite{MSW} Moody, Shapere and Wilczek 
have shown that nuclear motion in a diatomic molecule can be described by the effective non-Abelian  gauge field,  
\beq
{A}_i^{\;a}=(1-\kappa)\epsilon_{iaj}\,
\frac{x_j}{r^2}
\aand
\displaystyle{{F}^{\;a}_{ij}=(1-\kappa^2)\epsilon_{ijk}\frac{x_kx_a}{r^4}}\,,
\label{diatfields}
\eeq
respectively,
where $\kappa$ is a real parameter.
For $\kappa=0$, (\ref{diatfields}) is the field of the Wu-Yang monopole \cite{WuYang69}, \eqref{WYmonopA}.
For other values of $\kappa$, it is a truly non-Abelian configuration (except for
$\kappa=\pm1$, when the field strength vanishes and (\ref{diatfields}) is a gauge transform of the vacuum).
 
Dropping scalar potential $V(r)$  we return to  the Hamiltonian of a spinless particle with non-Abelian structure, \eqref{isomagham},
\beq
H = \half \vpi^2,\quad \vpi = \vp - \vA\,.
\label{mSWH}
\eeq

 Inquiring about  conserved quantities, we note first that 
when $\kappa\neq0$, then $q$ is not covariantly conserved in general,
\begin{equation}
D_jq=\frac{\kappa}{r}\left(Q^j-q\frac{x_j}{r}\right)
\neq 0\,,
\label{notpartr}
\end{equation}
implying that $q$ in \eqref{echarge} is \emph{not conserved} 
for $\kappa\neq0$, 
\beq
\quad
\big\{H,q\big\}=-\vpi\cdot \vD q \neq 0\,. 
\label{noecharge}
\eeq
unless the isospin is also radial. The bracketed quantity in \eqref{notpartr} is indeed the non-aligned-with-the-field piece of the isospin. When the isospin spin and the magnetic field happen to be aligned, then $q$ in \eqref{echarge} \emph{is} conserved.

Nor is the length of the to-become-charge $q$ is conserved in general,
\beq
\big\{H,q^2\big\}=-2\kappa q\,(\vpi\cdot\vD q)\neq0\,.
\eeq
whereas le the length of the isospin, $\vec{Q}^2$, \emph{is} conserved, 
$\{H,\vec{Q}^2\}=0$. Thus electric charge non-conservation comes from \emph{isospin precession}, as in the non-Abelian Aharonov-Bohm effect \cite{WuYang75,MSW,EZH-NABA}. For $\kappa=0$ we recover the Wu-Yang case when 
$q$ is conserved as we have seen in sec.\ref{WYmonopSec}.

The gauge field (\ref{diatfields}) is rotationally symmetric
and an isospin-carrying particle submitted to it
has, nevertheless conserved angular momentum \cite{MSW,Jackiw86}. Its form is, however, somewhat unconventional.

Our starting point is the first-order condition in (\ref{constraints}). We consider first $V=0$~; then  with 
$F_{jk}^a$  in (\ref{diatfields}), the equation to be solved is
\begin{equation}
D_iC=(1-\kappa^2)\,\frac{q}{r}\,\left((\vn\cdot\hx)\frac{x_i}{r}-n_i\right).
\label{diatC}
\end{equation}
In the Wu-Yang case, $\kappa=0$,  we have 
$C=- \vn\,\cdot\,q\,{\hx}$, but for $\kappa\neq0$ the to-be electric charge, $q$, is not conserved. Using (\ref{noecharge}) allows us to infer \cite{HP-NGOME} that
\begin{equation}
C 
= -\vn\cdot\Big(q\,{\hx}+\kappa(\vI -q\hx)\Big)\,.
\label{JC}
\end{equation}
The conserved angular momentum is, therefore,
\begin{eqnarray}
\vJ&=&\vx\times\vpi-\vW,
\\[6pt]
\vW&=&q\,{\hx}+\kappa\,(\vI -q\hx)=
q\,\hx+\kappa\,\big(\hx\times\vI\big)\times\hx,
\label{diatangmom}
\end{eqnarray}
consistently with the  results in \cite{MSW,Jackiw86}.  
Note however that the  spin-from-isospin contribution changes, w.r.t. \eqref{WYangmom},
\beq
q\,\hx\to\vW\,.
\eeq

For $\kappa=0$ we recover the Wu-Yang expression, (\ref{WYangmom}).
Eliminating $\vpi$ in favor of $\vp=\vpi+\vA$ allows us to rewrite the 
total angular momentum as
\beq 
\vJ=\vx\times\vp-\vI\, ,
\label{JpQ}
\eeq 
making manifest the ``spin from isospin term'' which is however \emph{not aligned with the ``hedgehog'' magnetic field. Consistently with \eqref{notpartr}, the non-conservation of $q$ in \eqref{echarge} comes precisely for this non-alignement}. 

Restoring the potential, we see that, again due to the non-conservation
of $q$, 
 $D_jV\neq0$ in general. The zeroth-order condition $\vC\cdot\vD V=0$ in (\ref{constraints}) is nevertheless satisfied if $V$ is a radial function which is independent of $\vI$,
$V=V(r)$, since then $\vD V=\vec{\nabla} V$, which is perpendicular to infinitesimal rotations, $\vC$. 
 Alternatively, a direct calculation, using the same formulae allow us to confirm that $\vJ$ 
commutes with the Hamiltonian, $\{J_i,H\}=0$. 

Multiplying (\ref{JpQ}) by $\hx$ yields by, once again,  
\beq
\vJ\cdot\hx=-q
\eeq  as in the Wu-Yang case.
This is, however, less useful as
before, since $q$ is \emph{not a constant of the motion anymore} so that the angle between  $\vJ$ and the radius vector, $\vx(t)$, is not constant either: the motion is not confined to a cone anymore.
  
Our attempts to find a conserved Runge-Lenz vector for the diatomic system have failed.

\section{Conclusion and outlook}\label{Concl}

The groundbreaking work of Kerner, \cite{Kerner68} and of Wong \cite{Wong70}, continued by many others, 
\cite{Trautman70,Cho75,Balachandran76,Balachandran77,Sternberg77,Sternberg78,Weinstein78,Sternberg80,DuvalCRAS,DuvalAix79,DH82,Montgomery,JackiwManton,FeherAPH} allow us to gain an insight into the structure of non-Abelian gauge theory \cite{YangMills}.
Our paper reviews the KK framework, retraces the chronological order of these discoveries and analyses the subtle physical meaning of isospin dynamics.

In addition to the conceptual works above, 
we underline that the Kerner-Wong model 
\cite{Kerner68,Wong70} admits important physical applications.

The analysis in a self-dual monopole field \cite{Feher86,Feher87,Feher:1988th, CFH90} could be paralleled by  studying motion in pure YM configurations with no scalar field \cite{Schechter,Wipf} and also  monopole scattering \cite{GibbonsManton86,FH86,Cordani88,CFH90,Feher:2009wwp}
alias Kaluza-Klein monopole \cite{GrossPerry,Cordani88,Sorkin}. 

In sec.\ref{ConsQuant} we  applied van Holten's algorithm  \cite{vHolten,vHolten2,vHolten3,HP-NGOME} conservation laws for a particle with isospin in  non-Abelian fields examplified by a Wu-Yang monopole \cite{WuYang69,vHolten}, and of diatomic molecules \cite{MSW,Jackiw86,HP-NGOME}. The O(4)/O(3,1) dynamical symmetry of a Wu-Yang monopole augmented with a self-dual Higgs field and implying elliptic trajectories is broken for diatomic molecules due to the non-conservation of the to-become electric charge, $q$ in \eqref{echarge}.

\vskip3mm
\begin{acknowledgments}\vskip-3mm This paper is dedicated to Richard Kerner 
 on the occasion of his 80th birthday. We are grateful to L. Feh\'er, M. Elbistan and L-P. Zou for correspondance and discussions. PMZ was partially supported by the National Natural Science Foundation of China (Grant No. 11975320). 
\end{acknowledgments}
\goodbreak
\


\end{document}